\documentclass[pdflatex,sn-mathphys]{HA-article}
\usepackage{physics}
\usepackage{pdfpages}

\jyear{2021}%

\begin{document}

\title[Hydrogenic Spin-Valley states of the Bromine donor in 2H-MoTe$_2$]{Hydrogenic Spin-Valley states of the Bromine donor in 2H-MoTe$_2$}


\author[1]{\fnm{Valeria} \sur{Sheina}}
\author[2]{\fnm{Guillaume} \sur{Lang}}
\author[3,4]{\fnm{Vasily} \sur{Stolyarov}}

\author[5]{\fnm{Vyacheslav} \sur{Marchenkov}}
\author[5]{\fnm{Sergey} \sur{Naumov}}
\author[5]{\fnm{Alexandra} \sur{Perevalova}}

\author[1]{\fnm{Jean-Christophe} \sur{Girard}}
\author[1]{\fnm{Guillemin} \sur{Rodary}}
\author[1]{\fnm{Christophe} \sur{David}}
\author[1]{\fnm{Leonnel Romuald} \sur{Sop}}
\author[1]{\fnm{Debora} \sur{Pierucci}}
\author[1]{\fnm{Abdelkarim} \sur{Ouerghi}}
\author[6]{\fnm{Jean-Louis} \sur{Cantin}}
\author[2]{\fnm{Brigitte} \sur{Leridon}}
\author[7]{\fnm{Mahdi} \sur{Ghorbani-Asl}}
\author[7,8]{\fnm{Arkady V.} \sur{Krasheninnikov}}
\author[1]{\fnm{Hervé} \sur{Aubin}}
\email{Herve.Aubin@universite-paris-saclay.fr}

\affil[1]{\orgdiv{Centre de Nanosciences et de Nanotechnologies (C2N)}, \orgname{UMR CNRS 9001}, \orgname{Université Paris-Saclay}, \orgaddress{\street{10 Boulevard Thomas Gobert}, \city{Palaiseau}, \postcode{91120}, \country{France}}}

\affil[2]{\orgdiv{Laboratoire de Physique et d'Étude des Matériaux}, \orgname{UMR CNRS 8213}, \orgname{ESPCI Paris}, \orgname{Université PSL}, \orgname{Sorbonne Université}, \orgaddress{\street{10 Rue Vauquelin}, \city{Paris}, \postcode{75005}, \country{France}}}

\affil[3]{\orgdiv{Advanced Mesoscience and Nanotechnology Centre}, \orgname{Moscow Institute of Physics and Technology}, \orgaddress{\city{Dolgoprudny}, \postcode{141700}, \country{Russia}}}

\affil[4]{\orgdiv{Advanced Mesoscience and Nanotechnology Centre}, \orgname{National University of Science and Technology MISIS}, \orgaddress{\city{Moscow}, \postcode{119049}, \country{Russia}}}

\affil[5]{\orgdiv{M.N. Mikheev Institute of Metal Physics}, \orgname{UB RAS}, \orgaddress{\city{Ekaterinburg}, \postcode{620108}, \country{Russia}}}

\affil[6]{\orgdiv{Institut des NanoSciences de Paris}, \orgname{UMR CNRS 7588}, \orgname{Sorbonne Université}, \orgaddress{\street{4 Place Jussieu}, \city{Paris}, \postcode{75005}, \country{France}}}

\affil[7]{\orgdiv{Institute of Ion Beam Physics and Materials Research}, \orgname{Helmholtz-Zentrum Dresden-Rossendorf}, \city{Dresden}, \postcode{01328}, \country{Germany}}

\affil[8]{\orgdiv{Department of Applied Physics}, \orgname{Aalto University, P.O. Box 11100}, \city{Aalto}, \postcode{00076}, \country{Finland}}

\abstract{In semiconductors, the identification of doping atomic elements allowing to encode a qubit within spin states is of intense interest for quantum technologies. In transition metal dichalcogenides semiconductors, the strong spin-orbit coupling produces locked spin-valley states with expected long coherence time. Here we study the substitutional Bromine Br\textsubscript{Te} dopant in 2H-MoTe$_2$. Electron spin resonance measurements show that this dopant carries a spin with long-lived nanoseconds coherence time. Using scanning tunneling spectroscopy, we find that the hydrogenic wavefunctions associated with the dopant levels have characteristics spatial modulations that result from their hybridization to the \textbf{Q}-valleys of the conduction band. From a Fourier analysis of the conductance maps, we find that the amplitude and phase of the Fourier components change with energy according to the different irreducible representations of the impurity-site point-group symmetry. These results demonstrate that a dopant can inherit the locked spin-valley properties of the semiconductor and so exhibit long spin-coherence time.}




\maketitle

\section{Introduction}



In zinc-blende III-V semiconductors, the large spin-orbit coupling leads to spin mixing and loss of spin coherence. This motivated the use of silicon, characterized by weak spin-orbit coupling, as a host of dopants for qubits \cite{Zwanenburg2013-ak,Salfi2014-ns,Usman2016-dh,Voisin2020-ar}. In contrast, in 2H-transition metal dichalcogenides (TMDCs), the combination of strong spin-orbit coupling and multiple valleys in the band structure provides protection against relaxation and decoherence. Indeed, due to the horizontal mirror symmetry $\sigma_h$ of the crystal structure, shown in Fig. 1a, the spin projection $s_z$ remains a good quantum number in the whole Brillouin zone, shown in Fig.~1b, even in the presence of strong spin-orbit coupling that lifts the degeneracy of the bands of opposite spin polarization. Furthermore, the electronic states are characterized by an additional index, the valley index, which is locked to the spin index \cite{Zhu2011-ss,Xiao2012-sz,Kosmider2013-jb,Liu2013-rm,Roldan2014-dz,Liu2015-ju,Kormanyos2015-bt}. This reduces the sensitivity of the spin to scattering processes because spin and valley indexes must change simultaneously. For this reason, numerous theoretical works have considered the spin-valley states hosted within 2H-TMDCs for encoding qubits \cite{Rohling2012-mx,Kormanyos2014-vt,Wu2016-ku,Pearce2017-xz,Szechenyi2018-jf,David2018-he, Pawlowski2018-xh}. Furthermore, it has been shown that Moir\'e lattices of spin-valley states could be used to simulate correlated \cite{Wu2018-ly,Schrade2019-mn} and topological \cite{Wu2019-qf} systems with recent experimental results obtained on twisted TMDCs bilayers\cite{Wang2020-pq,Tang2020-lo,Regan2020-ai,Li2021-yx,Li2021-qx}.

 
In monolayers, the valley of lowest (highest) energy in the conduction (valence) bands are located at the two nonequivalents \textbf{K} and $\mathbf{\bar{K}}$ points of the Brillouin zone. They are related by time reversal symmetry and the two Bloch states $\ket{\mathbf{K},\uparrow}$ and $\ket{\mathbf{\bar{K}},\downarrow}$ are Kramers partners capable of forming a qubit. To zeroth-order, spin-flip implies a change of valley and so the quantum states are protected from scattering by acoustic long-wavelength phonons \cite{Song2013-uo,Gilardoni2021-sv}. Time-resolved Kerr rotation measurements of the spin polarization of resident carriers have demonstrated spin lifetime reaching 100 ns for band electrons \cite{Yang2015-ww,Ersfeld2019-ol}, 4 to 40 ns for localized electrons \cite{Yang2015-hq,Jiang2022-ua} and 1~$\mu$s for holes \cite{Dey2017-tm}.


As the manipulation of one qubit requires the spin state to be localized spatially, this motivates the identification of dopants that inherit the locked spin-valley properties of the TMDC semiconductors. For this to happen, the dopant atomic orbitals must hybridize with the valley Bloch states and these Bloch states of opposite spin and valley polarization should not be mixed by the dopant confining potential. This means that each dopant quantum state should be formed from either the $\ket{\mathbf{K},\uparrow}$-valley or $\ket{\mathbf{\bar{K}},\downarrow}$-valley but not both. As detailed in Supplementary, group theory \cite{Evarestov1986-ct} shows that, for a dopant located on the anion site, the hybridization of the p-orbitals and the Bloch states at the \textbf{K}- and \textbf{Q}-valley is allowed by symmetry. Furthermore, it was shown \cite{Liu2014-vy,Wu2016-ku,Kaasbjerg2017-ta} that inter-valley mixing is forbidden by the C$_3$ symmetry of the anion site.

These conditions prompt the use of elements from column V (VII) as p (n) type dopants substituting the anion site. The formation of donor states near the conduction band by halogen dopants substituting the anion site has been confirmed by first principles calculations \cite{Komsa2012-md,Dolui2013-wx,Guo2017-xy,Onofrio2017-lz}, with Br\textsubscript{Te} having one of the lowest formation energies \cite{Onofrio2017-lz}. Furthermore, Br\textsubscript{Te} has been identified as an n-type dopant in earlier transport measurements \cite{Morsli1997-tu} and MoTe$_2$ has the largest spin-orbit coupling in the conduction band among Mo-based TMDCs.

In this work, we identify the Br\textsubscript{Te} spin signal by electron spin resonance (ESR) and relates the spin lifetime to the electronic properties obtained from transport measurements and angular resolved photoemission spectroscopy (ARPES). By scanning tunneling microscopy and spectroscopy (STM/STS), we demonstrate that the dopant levels are hydrogenic states hybridized to the \textbf{Q}-valleys of the conduction band. 





\section{Results}

The MoTe$_2$ crystal samples were grown by chemical vapor transport using Bromine gas as transport agent and doping source. To enable the preparation of Ultra-High Vacuum (UHV) clean surfaces by cleavage and because of insufficient sensitivity of standard ESR at measuring atomic monolayers of materials, we worked with bulk crystals. As discussed in Ref.~\cite{Zhang2014-ng}, while the global inversion symmetry is restored in bulk materials, the local inversion symmetry is still broken, meaning that the bulk material can be described as a stack of distinguishable layers. This picture has been confirmed by the observation of hidden spin-polarization of valleys \cite{Riley2014-hd,Razzoli2017-dv,Tu2020-zd} by spin-resolved ARPES and by measurements of valley orbital magnetic moment and Berry phase using circular-dichroism ARPES \cite{Cho2018-hu,Beaulieu2020-np}.

\subsection{ARPES and transport measurements}

Fig.~1c shows an ARPES spectra in $\mathbf{\Gamma}$-\textbf{K} direction measured on Br-doped MoTe$_2$. The band contours and the spin-orbit splitting of the valence band, $\approx 250$~meV, are consistent with DFT calculations \cite{Liu2013-rm}. At the $\mathbf{\Gamma}$-point, the valence band is about 1 eV below the Fermi energy, which implies that the bulk energy band-gap is about 1 eV as expected for MoTe$_2$ \cite{Liu2013-rm,Han2021-tr} and that the Fermi energy is in the conduction band. The transport properties are shown in Fig.~1d and Supplementary. From 300 K to 225  K, the resistivity decreases with temperature as expected in the saturation regime, where Hall measurements indicate a carrier concentration about $2 \times10^{18}$~cm$^{-3}$ and a Hall mobility reaching $\mu = 570~$cm$^2$V$^{-1}$s$^{-1}$, see Supplementary. From 100 K to 27 K, the resistivity follows an activated law with activation energy E$_\mathrm{A} =28~$meV, consistent with past works on Br-doped MoTe$_2$ \cite{Morsli1997-tu}. As will be confirmed by STM measurements shown below, this doping level can be described as a rescaled hydrogenic level n=1 with rescaled Bohr radius $\mathrm{a}_\mathrm{B} = \frac{\varepsilon_r}{\mathrm{m}^*}\mathrm{a}_0 \approx 2~$nm and Rydberg energy $\mathrm{E}_\mathrm{Ryd}  = 13.6\frac{\mathrm{m}^*}{\varepsilon_r^2} = 28~$meV, where we use for the dielectric constant $\varepsilon_r = 12 $\cite{Grasso1972-hi} and the effective mass $\mathrm{m}^* \approx 0.32$ \cite{Kormanyos2015-bt}. Below 27 K, the resistivity deviates from the activated law and enters an hopping regime, where the electrons are localized on the dopant and electronic transport occurs through tunnel hopping between the dopants. Below 15 K, the sample resistance is too large to be measurable with standard instruments. Following Ref.~\cite{Shklovskii1984-vr}, the temperature dependence of the resistivity is fitted by a Mott law $\rho \propto \exp(\xi_c)$ with the correlation length in two dimensions given by $\xi_c = (\mathrm{T}_0/\mathrm{T})^{1/3}$, which provides the temperature scale $\mathrm{T}_0\approx 27 \times 10^3$~K. From the correlation length, one obtains the average hopping length $\bar{r} = \mathrm{a}_\mathrm{B}\xi_c/4$. Using the Einstein relation between the mobility $\mu(T)=\mathrm{eD}/k_BT$ and the diffusion constant $D = \bar{r}^2/\tau_c$, the correlation time $\tau_c$, i.e., the delay between two hops, is obtained and shown in Fig. 1d. We will show now that this correlation time controls the spin lifetime measured by ESR.


\subsection{ESR measurements}

Fig.~1e shows the first-derivative ESR signal as a function of the amplitude of an in-plane magnetic field for different temperatures from 30 K down to 8 K. A resonance signal is visible only in the hopping regime, T $< 25~$K, of doped samples. No ESR signal has been observed in undoped samples obtained from HQ graphene. In a recent study\cite{Guguchia2018-gy} of undoped 2H-MoTe$_2$, while the signature of magnetism was observed from muon spin rotation measurements, no ESR signal could be observed. Figure~1f shows that the g-factor is anisotropic with $\mathrm{g}_\mathrm{zz}$ = 2.099 ($\mathrm{g}_\mathrm{xx}$ = 2.018) for the magnetic field perpendicular (parallel) to the sample plane. The anisotropy is opposite, $\mathrm{g}_\mathrm{zz}>\mathrm{g}_\mathrm{xx}$, and smaller than measured on arsenic acceptors in MoS$_2$ \cite{Title1973-sh,Stesmans2016-fq,Toledo2019-dh}. In TMDCs, a larger anisotropy for acceptors than donors is expected given the larger spin-orbit coupling in the valence band. Our value of $\mathrm{g}_\mathrm{zz}$ is consistent with DFT calculations of the spin contribution to the g-factor of localized electrons\cite{Kormanyos2014-vt,Pearce2017-xz} and Kerr measurements of the g-factor of localized electrons in monolayer MoS$_2$ \cite{Yang2015-hq,Jiang2022-ua}. The spectrum is constituted of a central line with additional sidelines and can be described by an effective spin Hamiltonian assuming two different contributions of identical g-tensor. The smallest contribution of weight 0.01 arises from electrons localized on single Br donors, i.e., not experiencing hopping, and produces the sidelines resulting from the hyperfine coupling of the electronic spin with the nuclear spin of the Br nucleus, where both natural isotopes have nuclear spin I = 3/2 for a total abundance of 100$~\%$. The second, largest, contribution of weight 0.99 produces the central line and arises from the donor electrons hopping between different Br sites, with the hyperfine structure being suppressed due to the different nuclear spin polarizations probed by the electron spin. A similar model was employed for arsenic acceptors in MoS$_2$ \cite{Title1973-sh}. An analysis of the angular dependence, shown in Supplementary, allows to extract the hyperfine and quadrupolar coupling constants and provides good fitting of the ESR data as shown in Fig.~1f. From the data measured as function of temperature, shown in Supplementary, we obtain the linewidth $\Delta \mathrm{B}_\mathrm{pp}$ as function of temperature, from which, the spin coherence \cite{Poole1996-zs} T$_2^* = \frac{2\hbar}{\sqrt{3}\mathrm{g}\mu_B\Delta \mathrm{B}_\mathrm{pp}}$ is calculated and shown in Fig.~1d, together with the correlation time obtained above. One clearly sees that at the highest temperature, T $\approx 25~$K, the spin lifetime is controlled by the correlation time, T$_2^* \approx \tau_c$, with no adjustable parameters. This indicates that strong Elliot-Yafet type dephasing occurs for each hop \cite{Meier2012-nr}. From this observation, we can conclude that the large spin-orbit coupling in MoTe$_2$ and Elliot-Yafet type dephasing is likely responsible for the disappearance of the resonance signal in the activated regime. Upon cooling into the hopping regime, the correlation time increases rapidly but the spin lifetime seems to saturate at a value T$_2^* \approx 5$~ns, which is similar to the spin lifetime of localized electrons in MoS$_2$ measured by Kerr rotation measurements\cite{Yang2015-hq}. The origin of this saturation remains to be understood, it could result from scattering with flexural phonons\cite{Song2013-uo} or exchange coupling between spins. As detailed in supplementary, the hyperfine-limited lifetime should be longer, about 100 ns. As already suggested in numerous previous works \cite{Rohling2012-mx,Song2013-uo,Kormanyos2014-vt,Wu2016-ku,Pearce2017-xz,Szechenyi2018-jf,David2018-he,Pawlowski2018-xh,Wang2018-sx,Gilardoni2021-sv}, the observation of spin lifetime larger than nanoseconds in TMDCs is likely the consequence of spin-valley locking. We show now STM measurements that indeed demonstrate that the bromine dopant levels are hybridized to the Bloch states of the \textbf{Q}-valleys.

\subsection{STM measurements}

Because the sample becomes insulating at liquid Helium temperature, T= 4.2 K, STM measurements are performed at liquid nitrogen temperature, T= 77 K. Based on previous STM works on undoped MoTe$_2$ \cite{Guguchia2018-gy} and MoSe$_2$ \cite{Edelberg2019-zq} as well as DFT calculations \cite{Gonzalez2016-rz}, we identified the molybdenum antisite Mo\textsubscript{Te}, shown in Fig.~2a, which has a characteristic hexagonal shape. In contrast, we see that the dopant Br\textsubscript{Te}, Fig.~2b, not observed in undoped samples, has a remarkable spatially modulated structure propagating over an area about 6 nm diameter centered on the original Te atomic site, see Supplementary.

From  several large scale topographic images, see Supplementary, the estimated surface density of Br dopants is n$_\mathrm{2D}\approx 4\times 10^{11}$ cm$^{-2}$ and corresponds to a bulk carrier density n$_\mathrm{3D} = n_\mathrm{2D}/t \approx 2.8\times 10^{18}$ cm$^{-3}$, where $t = 1.398$~nm is the length of unit cell along $z$, which is close to the Hall carrier density given above.

Fig.~2cd shows the two-dimensional fast Fourier transforms (2D-FFTs) of the topographic images. For the antisite Mo\textsubscript{Te}, only Bragg peaks are observed. For the dopant, Br\textsubscript{Te}, instead, the 2D-FFT shows peaks at wavevectors $\mathbf{m}_i = \mathbf{q}_j-\mathbf{q}_k$, $(i,j\neq k) \in \{1,2,3\}$, resulting from the interference between two \textbf{Q} valleys and peaks at wavevectors $\mathbf{q}_i=\mathbf{q}_j-\mathbf{\bar{q}}_k$, $(i,j\neq k) \in \{1,2,3\}$ and $\mathbf{h}_i=\mathbf{q}_i-\mathbf{\bar{q}}_i$, $i \in \{1,2,3\}$, resulting from the interference between \textbf{Q} and $\mathbf{\bar{Q}}$ valleys. 

Figure 3b shows the differential conductance $\dv{I}{V}{(V)}$, normalized by the integrated differential conductance, as function of sample bias (see methods). A comparison with the spectra measured on the pristine surface allows the identification of three energy ranges where the density of states (DOS) is modified by dopant levels, indicated as conduction band states (CBS) at bias $\approx 0.07$~V, in-gap states (IGS) at bias $\approx -0.7$~V and valence band states (VBS) at bias $\approx -0.9$~V. While a clear peak is only observed for the IGS, the CBS and VBS are merging with the bulk conduction and valence band states, so only shoulders are observed in the differential conductance. However, the CBS and VBS, as well as the IGS, can be clearly identified on Fig.~3c showing the differential conductance as a function of voltage and distance along a profile, indicated as a dashed line on the topographic image, Fig.~3a, running across the dopant. This plot shows that the DOS presents a spatial modulation on these three energy ranges. The conductance maps for the three energy ranges are shown in Fig.~3def. They present distinct spatial patterns but the corresponding 2D-FFTs, Fig.~3ghi, show peaks at the same wavevector coordinates $\mathbf{m}_i$, $\mathbf{q}_i$, $\mathbf{h}_i$ identified above. See Supplementary for additional maps at more energies. This modulation is not consistent with QuasiParticle Interferences (QPIs) of conduction electrons scattering on point-defects \cite{Roushan2009-zl,Liu2015-cn}. For QPIs, the interferences should be visible around all type of point-defects and the scattering wavevectors coordinates should depend on energy, following the Fermi surface contour. In this case, the \textbf{Q}-valleys interference would be visible only at the top of the conduction band, which is not what is observed experimentally.

Actually, these distinct spatial patterns result from a change of phase relationship between the Fourier components, as visible on the maps of the phase of 2D-FFTs, shown in Fig.~3jkl. Because the phase is not defined for complex numbers of zero amplitude, in these maps, the phase is shown only at k-vectors where the amplitude is large, within white circles. See Supplementary for details. To go further, we extract from the 2D-FFTs the amplitude and the phase of the Fourier components $\mathbf{m}_i$ as function of sample bias and plot them Fig.~3m and Fig.~3n, respectively. A peak in the Fourier amplitude is observed within the energy range corresponding to IGS but also for the CBS and VBS, which confirms that dopant-states are formed in these three energy ranges. Within each energy range, the phase remains nearly constant with values equal either a multiple of $\pi$ or a multiple of $\pi/3$. In-between, large phase jumps are observed and indicated by vertical red lines in Fig.~3n, at sample bias -0.015, -0.47 and -0.8 V. Similar behavior is observed for the components $\mathbf{q}_i$, shown in Supplementary.

As we will demonstrate now, the spatial modulation of the local DOS results from the hybridization of the dopant orbital levels to the Bloch states at the \textbf{Q}-valley and the phase-jumps are associated with changes of the symmetry of the eigenstates between the different energy levels. The formation of shallow hydrogenic dopant states in the multi-valley semiconductor silicon \cite{Kohn1955-pu,Ramdas1981-da} also leads to a spatial modulation of the DOS that has been observed only recently by STM \cite{Salfi2014-ns,Usman2016-dh,Voisin2020-ar}.

\subsection{Modeling of the dopant states}

As for both the Br and substituted Te atoms the valence states arise from their p-shell, the origin of the dopant levels can be figured out from simple arguments. In TMDCs, the d-orbitals of the Mo atom restrict to the irreducible representations (irreps) $\mathrm{A}_1'$, $\mathrm{E}'$ and $\mathrm{E}''$ of the D\textsubscript{3h} point group of the Mo site, $\mathrm{d}\downarrow\mathrm{D}_\mathrm{3h} = \mathrm{A}_1'\oplus\mathrm{E}'\oplus\mathrm{E}''$. Because these irreps are also induced by the p-orbitals of the Te atom,  $\mathrm{p}\uparrow\mathrm{D}_\mathrm{3h} = \mathrm{A}_1'\oplus\mathrm{A}_2'\oplus\mathrm{E}'\oplus\mathrm{E}''$, the d- and p-orbitals can hybridize and form bonds and bands. DFT calculations \cite{Liu2013-rm,Kormanyos2015-bt,Pike2017-na} show that the conduction band has $\mathrm{E}'$ symmetry and the valence band has $\mathrm{A}_1'$ symmetry. Both bands result from anti-bonding of Te p-orbitals and Mo d-orbitals as illustrated by the molecular diagram adapted from \cite{Pike2017-na} shown in Fig.~4a. The substitution of the Te atom with the Br atom will change the energy of the p-orbitals and affect both the $\mathrm{A}_1'$ and $\mathrm{E}'$ states. This implies that the CBS are likely formed from the $\mathrm{E}'$ originally located in the conduction band; the VBS and IGS are likely formed from the $\mathrm{A}_1'$, originally located in the valence band.

Using band representations theory, a theory of irreducible representations of space groups~\cite{Evarestov1986-ct}, detailed in supplementary, one can show that the p-orbitals of the Te (or Br) atoms restrict to the irreps $\mathrm{A}_1$ and $\mathrm{E}$ of the C\textsubscript{3v} point-group of the Te(or Br) site, $\mathrm{p}\downarrow\mathrm{C_\mathrm{3v}} = \mathrm{E}\oplus\mathrm{A}_1$, furthermore, one can show that the Bloch states at the \textbf{Q}-valleys restrict to the same irreps, $\mathrm{Q_{3(4)}}\downarrow\mathrm{C_\mathrm{3v}} = \mathrm{E}\oplus\mathrm{A}_1$. This decomposition is illustrated Fig.~4b. Consequently, the hybridization of the dopant p-orbitals and the \textbf{Q}-valley Bloch states are allowed to hybridize with symmetries $\mathrm{E}$ and $\mathrm{A}_1$. This is confirmed by DFT calculations \cite{Liu2013-rm,Liu2015-ju} showing that all three orbital components $p_{x,y,z}$ of the anion site have a large contribution to the \textbf{Q}-valleys in  the conduction band. Thus, two sets (representations) of dopant levels are expected: $\Gamma_{\text{CBS}}$ ($\Gamma_{\text{VBS}}$) resulting from the p-orbitals hybridized to the \textbf{Q}-valleys and located near the conduction (valence band).

As originally done for shallow dopants in silicon \cite{Kohn1955-pu,Ramdas1981-da}, the Br dopant quantum states are now described on the basis of the valley Bloch states. For each star (orbit) of wavevectors $\mathbf{Q}$ and $\mathbf{\bar{Q}}$, shown in Fig.~1b, there are three non-equivalent wavevectors. Furthermore, one star $\mathbf{Q}$ has opposite spin polarization to the other one $\mathbf{\bar{Q}}$. Thus, the states can be written as:

\begin{equation}
\begin{aligned}
\Psi_{\Gamma_i\mathbf{Q}} (\mathbf{r}, \downarrow)=&\frac{1}{\sqrt{3}}\mathrm{F}(\mathbf{r})\sum_j{\alpha_{ij}\phi_{\mathbf{q}_j\downarrow}(\mathbf{r}) }\\
\Psi_{\Gamma_i\mathbf{\bar{Q}}} (\mathbf{r}, \uparrow)=&\frac{1}{\sqrt{3}}\mathrm{F}(\mathbf{r})\sum_j{\alpha_{ij}^{\star}\phi_{\mathbf{\bar{q}}_j\uparrow}(\mathbf{r}) }\\
\end{aligned}
\label{eqn:psiBloch}
\end{equation}

with $\phi_{\mathbf{q}_js_z}(\textbf{r}) = u_{\mathbf{q}_j}(\mathbf{r})e^{i\mathbf{q}_j.\textbf{r}} \braket{\textbf{r}}{s_z}$ Bloch wavefunctions describing the valley states where $u_{\mathbf{q}_j}(\textbf{r})$ is the cell-periodic part and the envelope function $\mathrm{F}(\textbf{r}) \propto \exp(-\sqrt{x^2+y^2}/\mathrm{a}_\mathrm{B})$ describes the decay of the amplitude of the wavefunction with the Bohr radius calculated above, which describes properly the decay of the CBS as shown in Supplementary. The eigenstate $\Psi_{\Gamma_i\mathbf{Q}} (\mathbf{r}, s_z)$ should have the symmetry of the irrep $\Gamma_{i\mathbf{Q}}$ of the Br site symmetry point-group C\textsubscript{3v} where the index \textbf{Q} or $\mathbf{\bar{Q}}$ indicates from which valleys the state has been build of. Because the point-group C\textsubscript{3v} has only one and two-dimensional irreps, we expect the threefold valley degeneracy of the valley representations $\Gamma_{\text{CBS}}$ and $\Gamma_{\text{VBS}}$ to be lifted. In the case of silicon, this so-called valley-orbit splitting is of the order of few meV \cite{Zwanenburg2013-ak}. Following Kohn and Luttinger \cite{Kohn1955-pu,Ramdas1981-da}, we establish the characters of the valley representations to find that each one decomposes into one symmetric state A and one doublet state E, and this for each star $\mathbf{Q}$ or $\mathbf{\bar{Q}}$. Thus, each valley representation decomposes as $\Gamma_{\text{CBS(VBS)}} = $ A\textsubscript{1$\mathbf{Q}$} $\oplus$ E\textsubscript{$\mathbf{Q}$} $\oplus$ A\textsubscript{1$\mathbf{\bar{Q}}$} $\oplus$ E\textsubscript{$\mathbf{\bar{Q}}$}, as sketched in Fig.~4c. To describe spin-orbit coupling effects, we now use the double point-group C\textsubscript{3v} irreps obtained by taking the direct product of the simple point-group irreps with the spinor irrep E$_{1/2}$. We find that A generates the irrep E$_{1/2}$ and E generates the irreps E$_{1/2}$, E$_{3/2}$. As these irreps do not mix the Bloch states of the two different valleys \textbf{Q} and $\mathbf{\bar{Q}}$ and each valley has a well defined spin polarization, the dopant-states have well-defined spin and valley-polarizations, which implies spin-valley locking. As shown in Fig.~4c, the dopant-states formed from \textbf{Q}-valley must be spin-down and the dopant-states formed from $\mathbf{\bar{Q}}$-valley must be spin-up. Another set of dopant-states with reversed spin-polarization, not shown in Fig.4c, must exist but they are located at higher energy due to spin-orbit splitting. Assuming an energy separation of the order of the spin-splitting of the bulk states \cite{Liu2013-rm}, the two sets of dopant-states should be separated by about 250 meV for the VBS and IGS and about 15 meV for the CBS. For low temperature applications where the electron remains in the lowest energy state of the dopant, this second set of levels at higher energy can be ignored. For each irrep, using the standard operator projection method \cite{Dresselhaus2007-ed} and the character table of the double point-group C\textsubscript{3v}, the coefficients $\alpha_{ij}$ are calculated to give symmetry adapted linear combinations of the valley Bloch states. Then, the spatial distribution of the probability density $\rho(r) = \braket{ \Psi_i (\mathbf{r}, \sigma_i)}$ is calculated.

If no mixing occurs between the valleys \textbf{Q} and $\mathbf{\bar{Q}}$, the spatial pattern is composed of only the intra-valley Fourier components $\mathbf{m}_i$. This is illustrated Fig.~5afk for an irrep of A\textsubscript{1\textbf{Q}} symmetry. The spatial density map is shown Fig.~5a and the corresponding amplitude and phase of the 2D-FFT are shown Fig.~5f and Fig.~5k, respectively.

As sketched in Fig.~4d, the levels arising from the \textbf{Q}-valleys can be mixed with the levels arising from the $\mathbf{\bar{Q}}$-valleys in two distinct ways, either through changing the layer index with no spin-flip or, within the same layer, by changing the spin state.

For both cases, the resulting eigenstates can be written respectively as :
\begin{subequations}
\begin{align}
\Psi_{\Gamma_i} (\textbf{r}) = & \frac{1}{\sqrt{3}} \mathrm{F}(\mathbf{r}) \left[ \sum_{j}{\alpha_{ij}\phi_{\mathbf{q}_{j}\downarrow}(\mathbf{r}) } + e^{i\gamma} \sum_{j}{\alpha_{ij}\phi_{\mathbf{\bar{q}}_{j}\downarrow}(\mathbf{r}) } \right]  \\
\Psi_{\Gamma_i} (\textbf{r}) = & \frac{1}{\sqrt{3}} \mathrm{F}(\mathbf{r}) \left[ \sum_{j}{\alpha_{ij}\phi_{\mathbf{q}_{j}\downarrow}(\mathbf{r}) } + e^{i\gamma} \sum_{j}{\alpha_{ij}^{\star}\phi_{\mathbf{\bar{q}}_{j}\uparrow}(\mathbf{r}) } \right]
\end{align}
\end{subequations}

where $\gamma$ is an unknown phase factor that we take as 0 or $\pi$ to match the experimental data. Fig.~5b shows the spatial map in presence of valley mixing, using Eq.~2a to sum two states of symmetry A\textsubscript{1\textbf{Q}} and A\textsubscript{1${\mathbf{\bar{Q}}}$}. In addition to the intra-valley components $\mathbf{m}_i$, the Fourier map Fig.~5g shows additional components at the inter-valley components $\mathbf{q}_i$ and $\mathbf{h}_i$, as observed in experimental data.

Following the symmetry of eigenstates predicted by group theory, Fig. 4c, we now plot the corresponding local DOS.
Figure 5chm shows the spatial map of the probability density and the corresponding 2D-FFT resulting from the sum of four eigenstates, E\textsubscript{1/2\textbf{Q}} $\oplus$ E\textsubscript{3/2\textbf{Q}} $\oplus$ E\textsubscript{1/2${\mathbf{\bar{Q}}}$} $\oplus$ E\textsubscript{3/2${\mathbf{\bar{Q}}}$}, arising from $\Gamma_{\text{CBS}}$ and using Eq. 2a with $\gamma=\pi$. The result reproduces qualitatively the conductance maps of CBS, Fig.~3d. This comparison presumes that the STM spectroscopy measurements had not enough energy resolution to distinguish the states E\textsubscript{1/2} and E\textsubscript{3/2}. Figure 5din shows the results for the sum of four eigenstates, E\textsubscript{1/2\textbf{Q}} $\oplus$ E\textsubscript{3/2\textbf{Q}} $\oplus$ E\textsubscript{1/2${\mathbf{\bar{Q}}}$} $\oplus$ E\textsubscript{3/2${\mathbf{\bar{Q}}}$}, now arising from $\Gamma_{\text{VBS}}$ and using Eq. 2b with $\gamma=0$. The result reproduces qualitatively the conductance maps of IGS, Fig.~3e. In particular, the offset of the DOS maxima with respect to the image center indicated by plus symbol. Finally, Figure 5ejo shows the results for the sum of two eigenstates, E\textsubscript{1/2\textbf{Q}} $\oplus$ E\textsubscript{1/2${\mathbf{\bar{Q}}}$}, also arising from $\Gamma_{\text{VBS}}$ and using Eq. 2b with $\gamma=0$. The result reproduces qualitatively the conductance maps of VBS, Fig.~3f. A side by side comparison of the conductance maps and calculated probability densities for the three energy ranges are shown in Supplementary. 

Because of the large Bohr size, DFT calculations of the bulk Br-doped 2H-TMDC was too heavy, however, the calculation for one atomic monolayer is possible and shown in supplementary. A donor-state near the conductance band is identified and the corresponding local DOS presents a modulation resulting from the hybridization to the \textbf{K}-valleys, instead of the \textbf{Q}-valleys observed by STM in the bulk material. Furthermore, calculation of the partial density of states, Supplementary Fig.~16, shows that the $\mathrm{d}_{z^2}$ orbitals of the 1st and 2nd neighbors Mo atoms have a large contribution to this donor-state. As the Bloch states at the \textbf{K}-point of the conduction band have $\mathrm{d}_{z^2}$ character, this confirms the hybridization of the Br p-orbitals with the Bloch states at the \textbf{K}-valleys.

\section{Conclusion}

To summarize, we have identified the ESR signal of the Br\textsubscript{Te} dopant in 2H-MoTe$_2$ and found a spin state with long-lived (nanoseconds) coherence. This relatively long coherence time is believed to be the consequence of the protection by spin-valley locking. From STM measurements, we found that the dopant-orbitals are indeed hybridized to the \textbf{Q}-valleys. 


As discussed in Ref.~\cite{Gilardoni2021-sv}, in bulk materials, despite the local inversion symmetry, additional spin-scattering channels become possible. In particular, inter-layer coupling allows spin-flip without changing valley. Thus, we expect this work on bulk crystals to motivate STM and ESR studies of hydrogenic spin-valley states in doped atomic monolayer of TMDCs, where the protection afforded by spin-valley locking will reach its full potential. Our DFT calculations show indeed that the Br dopant levels in a single atomic monolayer are hybridized to the \textbf{K}-valleys.

\section{Figures}

\begin{figure*}[h]
\centering
 \includegraphics[width=0.7\textwidth]{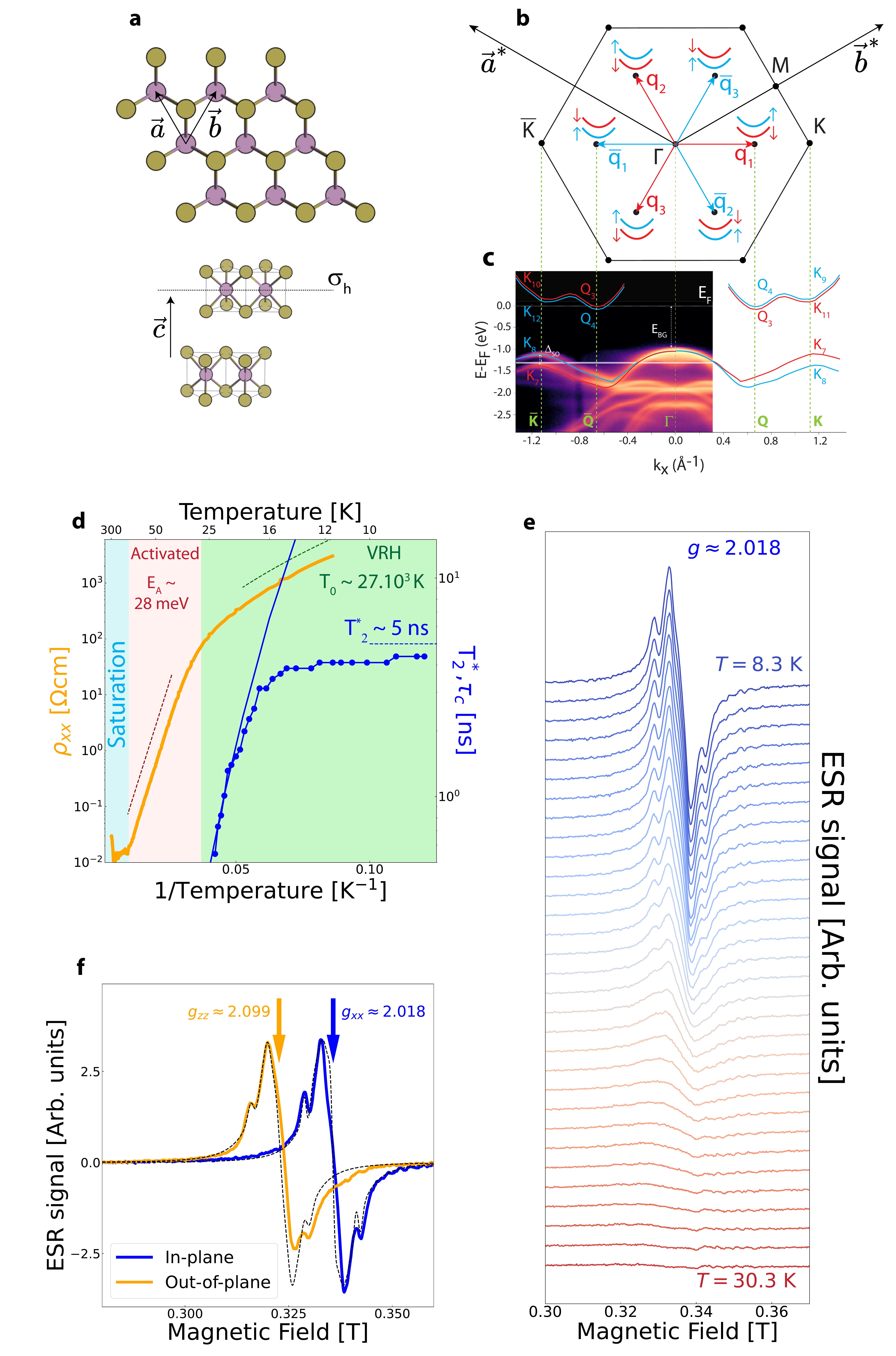}
	\caption{\textbf{Electronic properties of Br-doped 2H-MoTe$_2$.} \textbf{a}, Crystal structure of 2H-MoTe$_2$ where we indicate the basis vectors $\vec{a}$,$\vec{b}$,$\vec{c}$ of the Bravais lattice. The Mo atoms are in purple, the Te atoms in yellow. The horizontal mirror plane $\sigma_h$ is indicated as a dashed line. \textbf{b}, Brillouin zone where we indicate the basis vectors $\vec{a}^{\star}$,$\vec{b}^{\star}$ of the reciprocal lattice and the points of high symmetry. The star of wavevectors $\mathbf{Q}$ and $\mathbf{\bar{Q}}$ are indicated as red and blue arrows, respectively. \textbf{c}, Angular resolved photoemission (ARPES) map in the $\mathbf{\Gamma}$-\textbf{K} direction measured at T = 12 K. The continuous lines indicate the contours of the conduction bands (\textbf{Q}$_{3,4}$, \textbf{K}$_{9,10,11,12}$) and valence bands (\textbf{K}$_{7,8}$) extrema obtained from density functional theory (DFT) calculations \cite{Jain2013-hf}. The bands are labeled according to the irreps of the corresponding groups of wavevectors\cite{Song2013-uo,Gilardoni2021-sv}. The two colors, red and blue, indicate bands of opposite spin polarization. Note how the valleys of opposite momentum have opposite spin polarization. The \textbf{Q}-valleys are not visible in the ARPES data because of the low carrier density, however, the position of the Fermi level, indicated as an horizontal dashed line, 1 eV above the valence band, indicates the presence of the carriers in those \textbf{Q}-valleys. The band-gap E$_\mathrm{BG} \approx 1~$eV and the spin-orbit induced splitting $\Delta_\mathrm{SO} \approx 250~$meV are indicated. \textbf{d}, Arrhenius plot of resistivity (orange line), electron spin resonance (ESR) spin coherence lifetime T$_2^*$ (blue dot line) and hopping correlation time $\tau_c$ (blue continuous line) as function of temperature. The different transport regimes are identified by the colored panels: blue for the saturation regime, pink for the activated regime and green for the variable range hopping (VRH) regime. The dashed lines, displaced for clarity, are fits of the resistivity with the activated law and Mott law. \textbf{e}, ESR signal as function of temperature, from 30 K (red) down to 8 K (blue), measured with the magnetic field perpendicular to the c-axis, from which a g-factor $\mathrm{g} \approx 2.018$ is obtained. The ESR line is observed only below T~$\approx 25~$K, in the VRH regime. The spin coherence time T$_2^*$ extracted from the ESR linewidth is shown panel \textbf{d}. \textbf{f}, ESR signal measured for a magnetic field perpendicular (blue) and parallel (orange) to $\vec{c}$, from which the g-factor anisotropy is obtained. The dashed lines are fitted to the data using an effective spin Hamiltonian, see section "ESR measurements" and supplementary.}\label{fig1}
\end{figure*}

\begin{figure}[h]
\centering
   \includegraphics[width=1\textwidth]{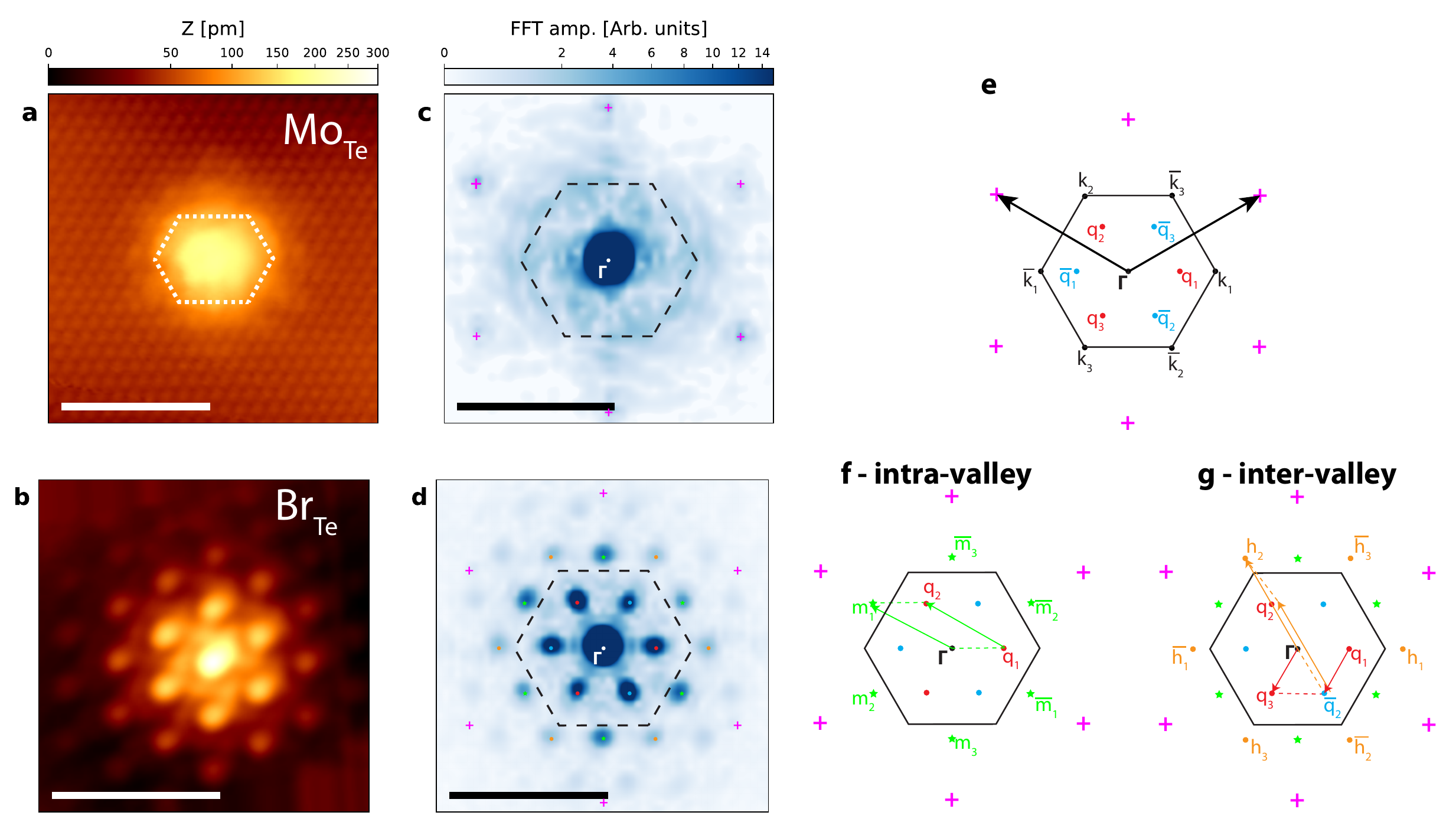}
	\caption{\textbf{Point-defects: Scanning tunneling microscopy (STM) topographies and two-dimensional fast Fourier transforms (2D-FFTs).} \textbf{ab},  STM topographies of Mo\textsubscript{Te} (I\textsubscript{setpoint} = 180 pA) and Br\textsubscript{Te}  (I\textsubscript{setpoint} = 400 pA), respectively, measured at sample bias -1 V and temperature of 77 K. The color bar quantifies the topographic height. A non linear color scale has been used to improve the visibility of Te atoms in the background. The white scale bar on each panel is 3 nm long. \textbf{cd}, Maps of the amplitude of the 2D-FFTs applied to the topographic images. The color bar quantifies the FFT amplitude. A non-linear color scale has been employed to improve the visibility of the FFT peaks of small amplitude. The black scale bar on each panel is equal to the length of the reciprocal lattice vector $\|\vec{a}^{\star}\| = 20.44$ nm$^{-1}$. \textbf{c,e}, For Mo\textsubscript{Te}, only the Bragg peaks (pink plus symbol) are observed. \textbf{efg}, For Br\textsubscript{Te}, peaks in the Fourier amplitude are observed at the intra-valley Fourier components $\mathbf{m}_i = \mathbf{q}_j -\mathbf{q}_i$ (green star symbols) and peaks of strongest amplitude are observed at the inter-valley Fourier components $\mathbf{q}_i = \mathbf{q}_j - \mathbf{\bar{q}}_i$ (red and blue disc symbols) and $\mathbf{h}_i = \mathbf{q}_i - \mathbf{\bar{q}}_i$ (orange disc symbols). The arrows show how the Fourier components arise from the valleys wavevectors $\mathbf{q}_j$ and $\mathbf{\bar{q}}_i$.
    }\label{fig2}
\end{figure}

\begin{figure}[h!]
	\centering
	  \includegraphics[width=1.1\textwidth]{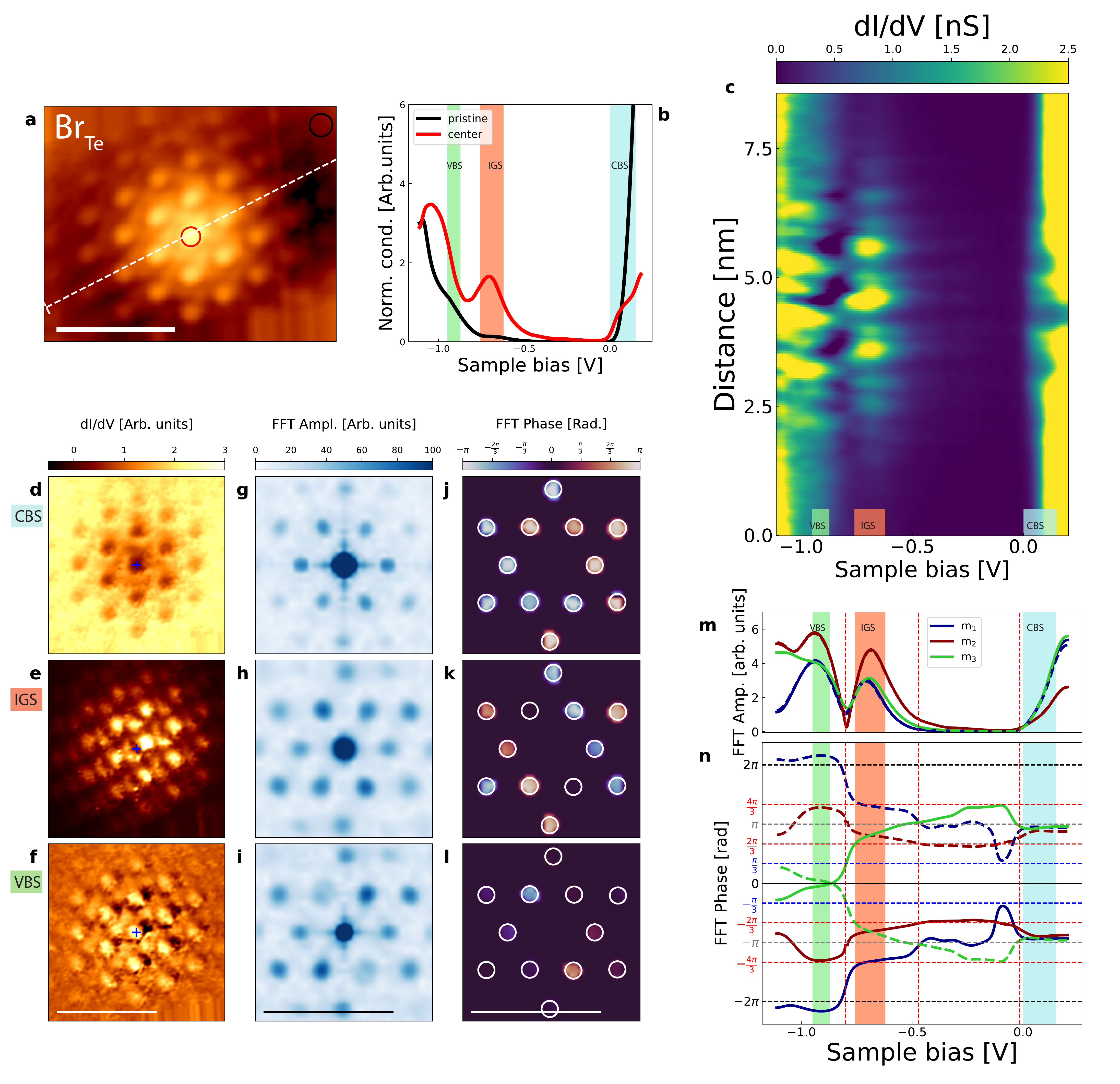}
	\caption{\textbf{Br\textsubscript{Te} : Scanning tunneling spectroscopy (STS), conductance maps and two-dimensional fast Fourier transforms (2D-FFTs).} \textbf{a}, Scanning tunneling microscopy (STM) topography of Br\textsubscript{Te}. \textbf{b}, Differential conductance $\dv{I}{V}{(V)}$ as function of sample bias. The red (black) curve is obtained by averaging the spectra within the circle located at the center (away) from the dopant, shown on panel \textbf{a}. The ranges corresponding to valence band states (VBS), in-gap states (IGS) and conduction band states (CBS) are indicated as green, red and blue zones, respectively. \textbf{c}, Differential conductance map as function of sample bias and distance along a path going through Br\textsubscript{Te} center, shown as a white dashed line in panel \textbf{a}. The color bar quantifies the conductance value. The origin of the distance scale starts at the most left end of the white dashed line. The VBS, IGS and CBS ranges are indicated at the bottom. \textbf{def}, Differential conductance maps at sample bias of 0.07, -0.7 et -0.9 V, corresponding to the CBS, IGS and VBS, respectively. The scale bar shown on panel \textbf{f} is 3 nm long. As a guide to eye, a plus symbol indicates the center of the image. \textbf{ghi}, Maps of the amplitude of 2D-FFTs applied to the conductance maps. Note that the wavevectors coordinates of the maxima are not changing with energy. \textbf{jkl}, Maps of the phase of the 2D-FFTs. Note that the phase pattern is changing with energy. The scale bars shown on panels \textbf{il} are equal to the length of the reciprocal lattice vector $\|\vec{a}^{\star}\|$. \textbf{m}, Plot of the amplitude of the Fourier components $\mathbf{m}_i$ (continuous line) and  $\mathbf{\bar{m}}_i$ (dashed line) as function of sample bias. Note that the amplitude is large only in the colored zones corresponding to the VBS, IGS and CBS. \textbf{n}, Plot of the corresponding phase for the same components. Note that the phase remains constant in the energy ranges VBS, IGS and CBS, where the phase value is either a multiple of $\pi$ or a multiple of $\pi/3$. Note the rapid phase shift, indicated by vertical red lines, at sample bias -0.015 V, -0.47 V and -0.8 V, separating the CBS from the IGS from the VBS, successively.}
	\label{fig3}
\end{figure}

\begin{figure}[h!]
	\centering
	  \includegraphics[width=1\textwidth]{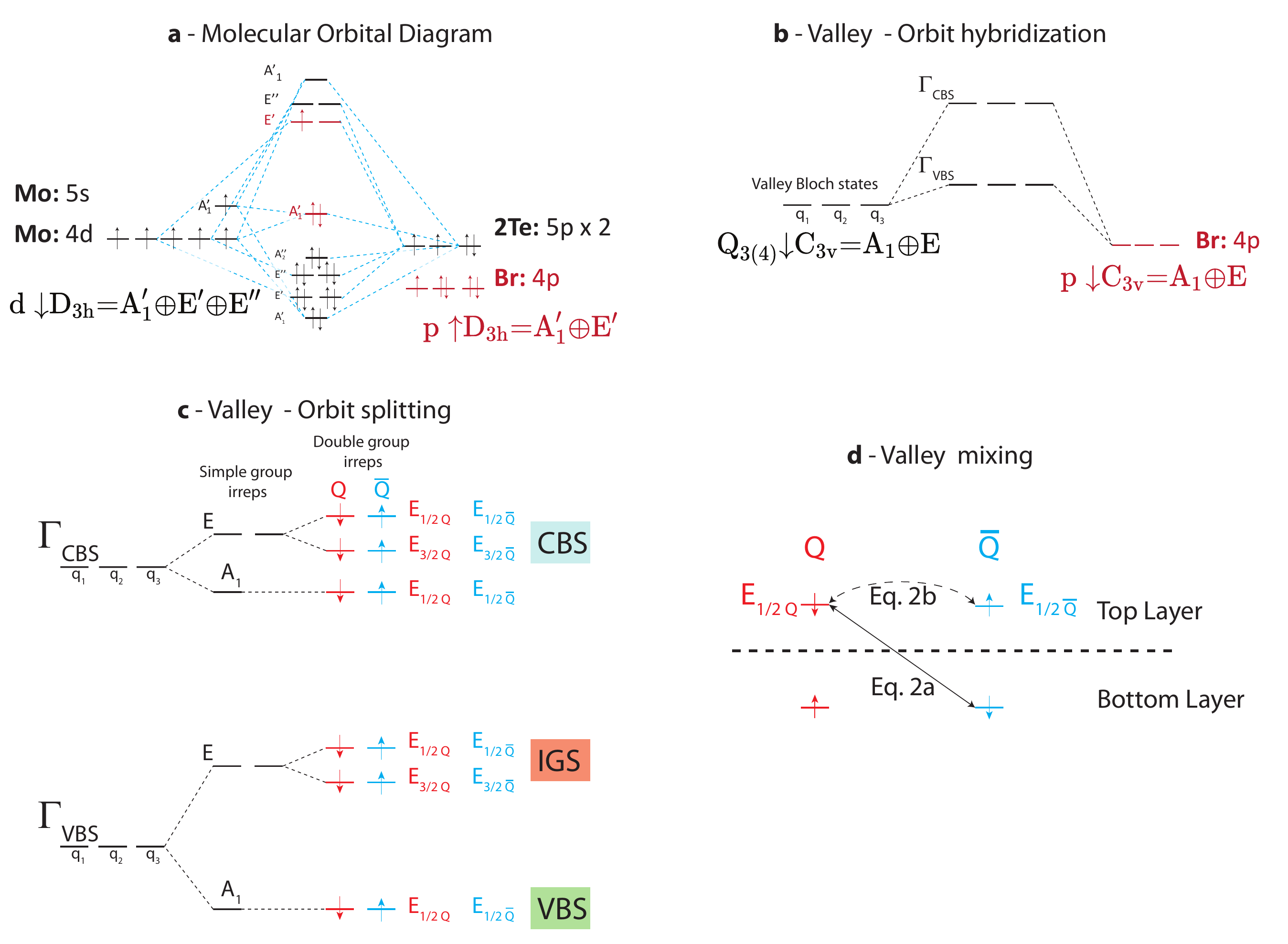}
	\caption{\textbf{Modeling of dopant levels.} \textbf{a}, Molecular orbital diagram adapted from Ref.~\cite{Pike2017-na}, indicating the hybridization of the orbitals according to irreps of the point-group D\textsubscript{3h}. Upon substituting the Te atom with a Br atom, the p-orbitals shift down in energy, leading to conduction band states (CBS) of E' symmetry, originally located in the conduction band, and in-gap states (IGS), valence band states (VBS), of A'$_1$ symmetry, originally located in the valence band. \textbf{b}, Group theory shows that the hybridization of the Bloch states of the \textbf{Q}-valleys and the p-orbitals of the anion site is allowed by symmetry. This leads to two valley representations of dopant levels, one located near the valence band, $\Gamma_{\text{VBS}}$ and one located near the conduction band $\Gamma_{\text{CBS}}$. Two additional valley representations are formed from the Kramers partners at $\mathbf{\bar{Q}}$. \textbf{c}, Each valley representation, $\Gamma_{\text{VBS}}$ and $\Gamma_{\text{CBS}}$, splits into irreps A$_1$ and E of the simple point-group C\textsubscript{3v}, which gives three irreps (2$\times$E$_{1/2}$, E$_{3/2}$) of the point double-group C\textsubscript{3v}. \textbf{d}, Illustration of inter-valley mixing either through interlayer coupling without spin flip as described by Eq. 2a or through inter-valley coupling with spin-flip as described by Eq. 2b.}
	\label{fig4}
\end{figure}

\begin{figure}[h!]
	\centering
	 \includegraphics[width=0.9\textwidth]{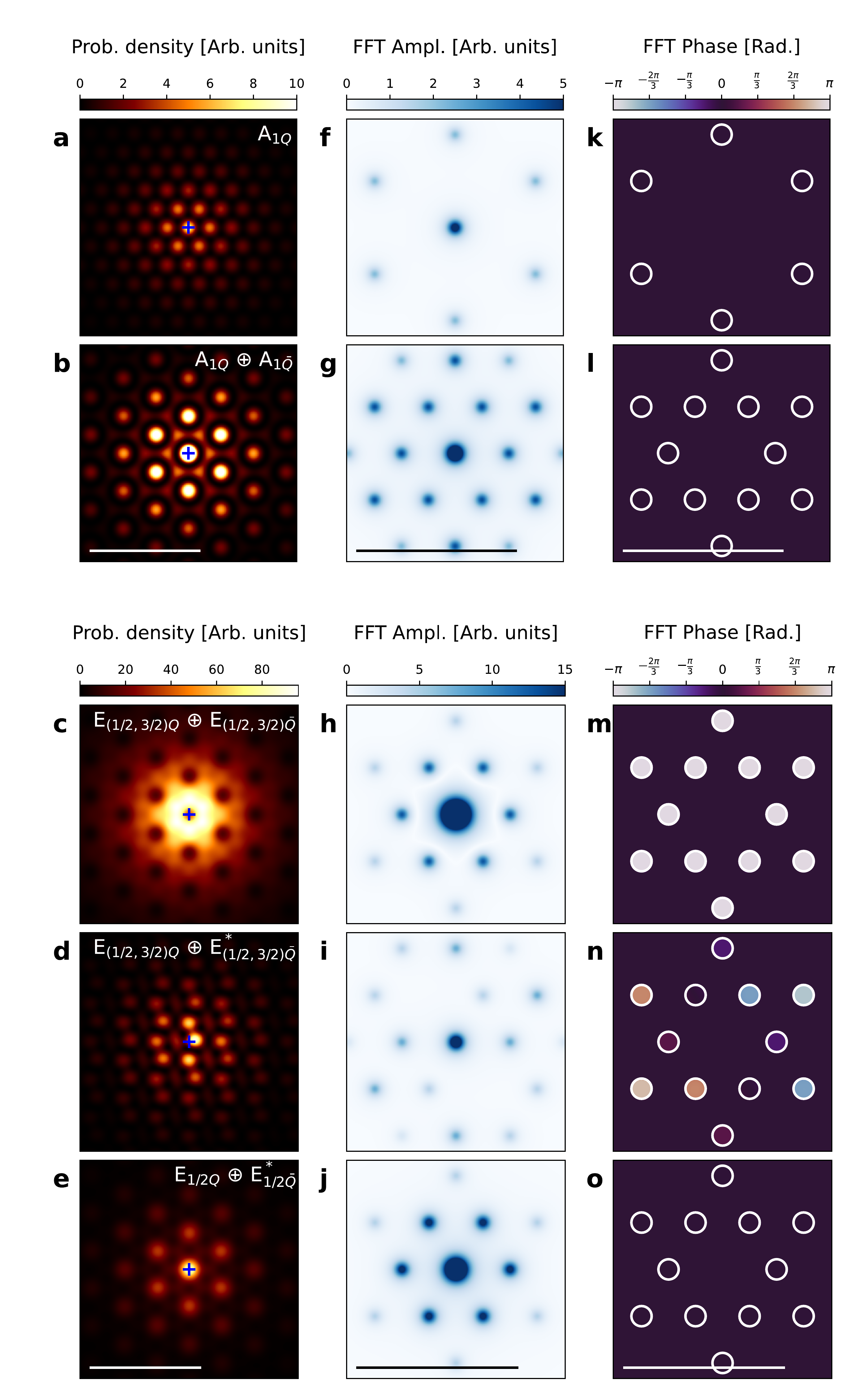}
	\caption{\textbf{Symmetry adapted linear combination of Bloch valley states.} \textbf{a-e}, Maps of the probability density for dopant states transforming as different irreps or combination of different irreps as indicated in the panels. See section "Modeling of the dopant states" for details. The color-bar quantifies the amplitude. The white scale bars are 3 nm long. On each panel, a plus symbol indicates the center of the image. \textbf{f-j}, Maps of the amplitude of the two-dimensional fast Fourier transforms (2D-FFTs) applied to the probability density maps. The color-bar quantifies the amplitude. The black scale bars are equals to the length of the reciprocal lattice vector $\|\vec{a}^{\star}\|$. \textbf{k-o}, Maps of the phase of the 2D-FFTs applied to the probability density maps. The color-bar quantifies the value of the phase. The white scale bars are equals to the length of the reciprocal lattice vector $\|\vec{a}^{\star}\|$.}
	\label{fig5}
\end{figure}
\clearpage

\section{Methods}
\subsection{Crystal growth}
MoTe$_2$ single crystals were grown by the chemical vapor transport (CVT) method using Br$_2$ as a transport agent \cite{Levy1977-nf}. Stoichiometric mixture of molybdenum and tellurium powders along with Br$_2$ were sealed in a quartz ampoule with a length of 24 cm and diameter of 1.5 cm. The bromine vapor density was approximately 5 mg/cm$^3$. The ampoule was pumped out to a residual pressure of $\approx$ 10$^{-4}$ atm. and then placed into a horizontal tube furnace with a linear temperature gradient. The temperatures of the hot zone T$_1$ and the cold zone T$_2$ were 850 $^{\circ}$C and 770 $^\circ$C, respectively. In the hot zone, the precursor reacted with the gaseous transport agent to form volatile compounds, which, under the action of a temperature gradient, were transferred to the opposite end of the ampoule (cold zone), where MoTe$_2$ single crystals grew and Br$_2$ was released. The single crystal growth procedure was carried out for 500 hours, followed by slow cooling to room temperature. The crystalline structure was checked by X-ray diffraction where we found that MoTe$_2$ crystallized in a hexagonal structure (Space group P63/mmc (\#194)) with the lattice parameters a = 3.540(7) \AA~and c = 13.983(5) \AA.

\subsection{Transport measurements}
Transport measurements were carried out in a Physical Property Measurement System (PPMS). The longitudinal and Hall resistance were measured using a standard lock-in technique. For these measurements, the bulk crystals were exfoliated down to obtain thin crystals about ten micrometers thick, deposited on an insulating silicon wafer. The electrical contacts were realized with gold wires ($\varnothing 25 \mu m$) glued with silver epoxy.

\subsection{Photoemission}
The ARPES measurements were conducted at the CASSIOPEE beamline of Synchrotron SOLEIL (France). Before the measurement, the sample was cooled down at T $\approx$ 12 K and cleaved in situ in UHV in the analysis chamber. The CASSIOPEE beamline is equipped with a Scienta R4000 hemisperical electron analyzer (angular acceptance ±15$^{\circ}$) with vertical slits. The angular and energy resolutions were 0.25$^\circ$ and 15 meV. The mean diameter of the incident photon beam was smaller than 50 µm. We used linear horizontal polarized photons with an energy of 47 eV. Binding energies are referenced to the Fermi level of a gold foil in electrical contact with the sample.

\subsection{ESR measurements}
The samples were studied with two Bruker spectrometer, EMX and ELEXYS-II, working in CW-mode in X-band in a cavity of frequency 9.482 GHz. The thin flat bromine-doped 2H-MoTe$_2$ crystal was glued on the flat part machined into a glass rod, enabling the control of the angular orientation of the sample with respect to the applied magnetic field. The angle is measured with respect to the axis perpendicular to the sample. The orientation was manually controlled with a goniometer of 0.5 deg precision. The spectrometer provides the first derivative of the absorbed microwave power as a function of magnetic field. Measurements were carried out in a helium-flow cryostat in the temperature range from T = 4 K to T = 300 K. Changing the in-plane orientation did not lead to significant changes in the spectrum and so the in-plane orientation has not been determined.

\subsection{STM measurements}
The bulk 2H-MoTe$_2$ doped crystals were cleaved under UHV conditions, P $< 2.10^{-10}$ mbar, to get clean surfaces free of atomic contamination. The samples were measured at T = 77 K in two different microscopes: one Joule-Thomson (JT) from SPECS and one LT from Omicron (Scienta). The differential conductance $\dv{I}{V}{(V)}$ spectra are taken with the feedback loop open with current setpoint set at sample bias of -1.2 V. To compare spectra measured at different locations or plotting conductance maps, we need to remove the effects of changing tunnel barrier height. To that end, we assume that the total DOS is conserved on the energy range [-1.2 V, 0.15 V ]. Thus, we normalize all measured $\dv{I}{V}{(V)}$ curves by their integrated values $\int_{-1.2}^{0.15}{\dv{I}{V}{}dV}$.

\backmatter

\section*{Availability of data and materials}
Any further data are available from the corresponding author upon request.

\section*{Code availability}
Most of data analysis and plotting were done under python, expect for the analysis of ESR data which were done with Matlab code (Easy spin). All codes are available upon request.

\section*{Acknowledgments}
We acknowledge financial support from ANR MECHASPIN Grant No. ANR-17-CE24-0024-02 and ANR FRONTAL Grant No. ANR-19-CE09-0017-02. We acknowledge support from the CNRS research infrastructure RENARD (FR 3443) for EPR facilities. The crystal growth was carried out within the state assignment of Ministry of Science and Higher Education of the Russian Federation (theme “Spin” No. 122021000036-3). The computational support from the Technical University of Dresden computing cluster (TAURUS), from High Performance Computing Center (HLRS) in Stuttgart, Germany is gratefully appreciated.
We acknowledge useful discussions regarding samples with Dr. B. Fauqué. We thank Pr. H. Dery, Dr. CM. Gilardoni and Pr. M. Guimaraes for careful reading of the manuscript and suggestions.

\section*{Authors' contributions}
H.A. proposed the project. V.M., S.M and A.P. grew the MoTe$_2$ crystal. V.Sh., G.L. H.A. and J.L.C. realized the ESR measurements. V.Sh., H.A. and B.L. realized the transport measurements. D.P and A.O. realized the ARPES measurements. V.Sh., V.St., H.A., J.C.G, G.R., C.D., L.R.S., realized the STM measurements. M.G.A. and A.V.K. performed the DFT calculations. V.Sh. and H.A. realized the group theory analysis and wrote the paper with contributions of all authors.

\section*{Conflict of interest/Competing interests}
The authors declare no competing interests.

\section*{Declarations}

\bmhead{Supplementary information}
The online version contains supplementary material  available at ..
\bmhead{Correspondence and requests for materials} should be addressed to H.A.

\bibliography{biblio}

\includepdf[pages=-]{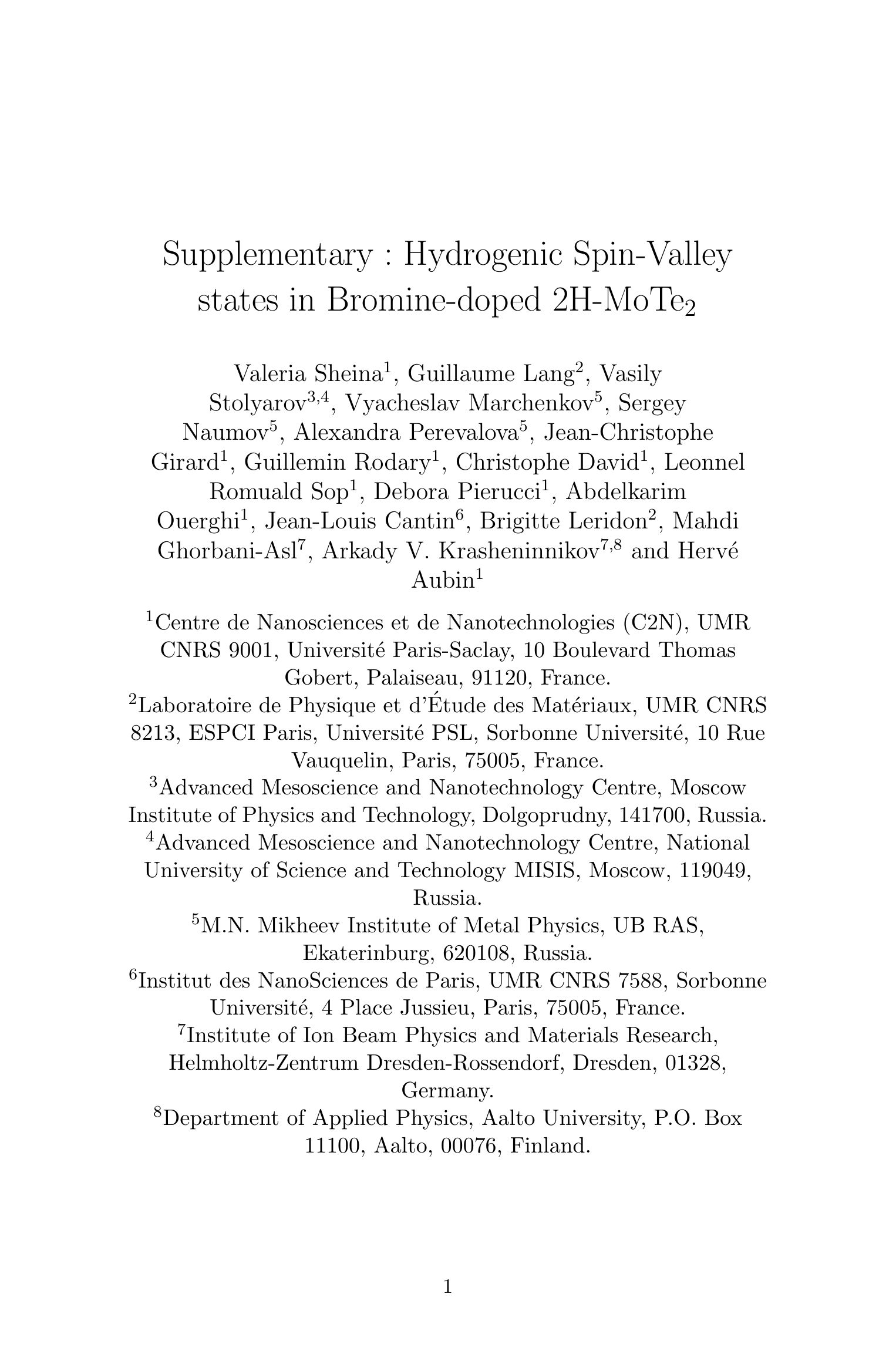}
\end{document}